# Reduction in Thermal Conductivity of Monolayer MoS$_2$ by Large Mechanical Strains for Efficient Thermal Management

*Jun Liu[1], Mengqi Fang[1], Eui-Hyeok Yang[1], Xian Zhang[1]\**

[1]Department of Mechanical Engineering, Stevens Institute of Technology, Hoboken, New Jersey 07030, United States

KEYWORDS. molybdenum disulfide, thermal conductivity, interfacial thermal conductance, large mechanical strains, Raman spectroscopy.

ABSTRACT. Two-dimensional (2D) materials such as graphene and transition metal dichalcogenides (TMDC) have received extensive research interests and investigations in the past decade. In this research, we report the first experimental measurement of the in-plane thermal conductivity of $MoS_2$ monolayer under a large mechanical strain using optothermal Raman technique. This measurement technique is direct without additional processing to the material, and $MoS_2$'s absorption coefficient is discovered during the measurement process to further increase this technique's precision. Tunable uniaxial tensile strains are applied on the $MoS_2$ monolayer by stretching a flexible substrate it sits on. Experimental results demonstrate that, the thermal conductivity is substantially suppressed by tensile strains: under the tensile strain of 6.3%, the thermal conductivity of the $MoS_2$ monolayer drops approximately by 62%. A serious of thermal transport properties at a group of mechanical strains are also reported, presenting a strain-dependent trend. It is the first and original study of 2D materials' thermal transport properties under a large mechanical strain (>1%), and provides important information that the thermal transport of $MoS_2$ will significantly decrease at a large mechanical strain. This finding provides the key information for flexible and wearable electronics thermal management and designs.

TEXT.

INTRODUCTION

Two-dimensional (2D) monolayer molybdenum disulfide ($MoS_2$) has stimulated intensive research due to its intriguing physical properties such as large surface-to-volume ratio, inherent flexibility, and sizable bandgaps[1-2]. Such properties potentially enable the development of flexible and wearable thermoelectric devices, which are designed to directly convert body heat into electricity according to the Seebeck effect[3-5]. The energy conversion efficiency of such devices can be determined by the dimensionless figure of merit, $ZT = S^2\sigma T/\kappa$, where $S$ is the Seeback coefficient, $\sigma$ is the electrical conductivity, $T$ is the temperature and $\kappa$ is the thermal conductivity. It is indicated that the thermal conductivity of $MoS_2$ can tune the figure of merit and determine the performance of the flexible/wearable thermoelectric devices[6]. For the operation of such devices, $MoS_2$ usually experiences mechanical deformations to achieve intimate conformal contact with human skin and to coordinate complex human motions. Understanding how the thermal conductivity of $MoS_2$ changes with mechanical deformation thus is essential.

While there are regular experimental and theoretical investigations on the thermal conductivity of $MoS_2$[7-13], the effect of mechanical strains on the thermal conductivity is still rarely studied. So far only a few simulation predications have been conducted and the conclusions are quite diverse. Jiang and co-workers conduct molecular dynamics (MD) simulations to investigate the strain effect on the thermal conductivity of an armchair $MoS_2$ monolayer at 300 K[14]. It is found that the thermal conductivity of $MoS_2$ drops by around 40% at the tensile strain of 8%. Ding and co-workers also report that the thermal conductivity of monolayer $MoS_2$ is heavily suppressed by uniaxial tensile strains: under the tensile strain of 12%, the thermal conductivity of the $MoS_2$ monolayer decreases approximately 70%[15]. Using first-principles calculations, Xiang and co-workers find that the thermal conductivity of $MoS_2$ monolayer almost linearly decreases with the uniaxial tensile strain[16]. The reduction is attributed to the strain-induced scattering of the acoustic phonon modes, which leads to a decrease in the group velocity, specific heat, and vibrational transmission. Zhu and co-workers investigate the effect of biaxial tensile strain on the thermal conductivity of $MoS_2$ by combining first-principles calculations and the Boltzmann transport equation[17]. It is reported that a moderate biaxial tensile strain of 2~4% results in a 10~20% reduction in the thermal conductivity. They emphasize that the reduction in the thermal conductivity is mainly attributed to the low-frequency phonons because high-frequency phonons have much smaller group velocities, suffer intensive phonon-phonon scattering, and possess short relaxation times as compared with that of the low-frequency phonons. As a result, the contribution of the high-frequency phonons to the thermal conductivity is negligible. Wang and Tabarraei use nonequilibrium MD simulation to study the effect of uniaxial stretching on the thermal conductivity of $MoS_2$ nanoribbons[18]. Their results, however, demonstrate that the thermal conductivity is not sensitive to the tensile strains. Zhang and co-workers investigate the strain effect on the thermal conductivity of $MoS_2$ via MD simulations and also claim that the uniaxial tensile strain has weak effects on the thermal conductivity[19]. The large disparity in thermal conductivities predicted by the previous computational work motivates further experimental investigation. It is admitted that the experimental realization and characterization of such small material remains a big challenge, while it is still quite necessary to design and conduct measurements on the thermal properties of $MoS_2$ to address the controversy, which may shed some light on designing wearable/flexible thermoelectric devices with high efficiency.

A systematic experimental study on the effect of mechanical strain on the in-plane thermal conductivity of MoS$_2$ is still missing so far. To fill this gap, we perform an opto-thermal Raman characterization method to study the dependence of the in-plane thermal conductivity of a MoS$_2$ monolayer on uniaxial tensile strain. Controlled uniaxial tensile strains are purposely introduced to the MoS$_2$ monolayer by stretching a deformable substrate (polydimethylsiloxane, PDMS) it sits on. The first-order temperature coefficient and the laser power-dependent Raman peak shift rates are recorded by monitoring the red shift of the Raman peak of MoS$_2$. The in-plane thermal conductivity of the MoS$_2$ monolayer under different tensile strains is determined through solving the heat diffusion equation in cylindrical coordinates. The thermal conductivity of the MoS$_2$ monolayer is found to decrease significantly with the uniaxial tensile strains.

RESULTS AND DISCUSSION

Opto-thermal Raman technique has been widely applied to measure the thermal conductivity of graphene and 2D MoS$_2$ due to its non-destructive and contactless natures[20-22]. The temperature distribution, $\kappa$, and $g$ of the sample are described by the heat diffusion equation in cylindrical coordinates in the following equation.

$$\frac{1}{r}\frac{d}{dr}\left(r\frac{dT}{dr}\right) - \frac{g}{\kappa t}(T - T_a) + \frac{Q}{\kappa} = 0 \qquad (1)$$

where $t$ is the thickness of the sample, $T_a$ is the global temperature, and $T$ is the temperature in the sample at position $r$ upon laser heating. Assuming a Gaussian beam profile, the volumetric optical heating, $Q$, is expressed by the following equation.

$$Q = \int_0^\infty q_0 e^{(-r^2/r_0^2)} 2\pi r \, dr = q_0 \pi r_0^2 \qquad (2)$$

where $q_0$ is the peak of the absorbed laser power per unit area at the center of the beam spot, $r_0$ is the radius of the laser spot.

To extract the two unknowns of $\kappa$ and $g$, the rise of the temperature ($T_m$) in the sample measured by the Raman laser beam is determined by the local temperature distribution, weighted by the Gaussian profile of the laser spot in the following equation.

$$T_m = \frac{\int_0^\infty T e^{(-r^2/r_0^2)} r dr}{\int_0^\infty e^{(-r^2/r_0^2)} r dr} \tag{3}$$

This temperature is correlated to the measured thermal resistance defined by the equation of $R_m = T_m/Q$ and $R_m$ can be determined by measuring the first-order temperature coefficient, $\chi_T$, and the laser power-dependent Raman peak shift rate, $\chi_P$. For measuring $\chi_T$, the temperature of the sample is externally tuned from 300K to 390K using a heater placed at the bottom of the PDMS substrate. For measuring $\chi_P$, the Raman spectrometer is employed as the heat source for producing a local temperature rise and to characterize the Raman frequency shift induced by the laser beam. Two laser power-dependent Raman peak shift rates are measured using both 50× and 100× objective lenses for extracting the two unknowns of $\kappa$ and $g$.

The laser beam size of each objective is obtained by scanning across a sharp edge of a bulk $MoS_2$ on a $SiO_2$/Si substrate and measuring the intensity of the silicon peak at around 520 cm$^{-1}$ as a function of the distance from the edge. Through fitting the experimental data to a Gaussian error function, the determined beam sizes for 50× and 100× objective lenses are 0.34 μm and 0.25 μm, respectively. More details can be found in the supplementary materials of Fig. S1.

The incident laser power for all measurements is lower than 550 μW to avoid possible damage to the sample and to stay within the liner dependence range. The absorbed laser power, $q_0$, is calculated as $q_0 = \alpha q$, where $q$ is the incident laser power measured using a power meter, $\alpha =$ 0.045 is the absorbance of the chemical vapor deposition (CVD) MoS$_2$ monolayer[23]. All Raman spectra in this study are collected by using a confocal Raman microscopy system with the excitation laser wavelength of 514 nm.

Controlled tensile strains are purposely introduced to the MoS$_2$ monolayer by stretching a deformable substrate (polydimethylsiloxane, PDMS) it sits on. To apply uniaxial tensile strain, the PDMS substrate is mounted on a home-made testing machine. This fabrication method is similar in another work on the optical properties of MoS$_2$ under strains which is using a flexible polyethylene terephthalate (PET) substrate[24]. The mechanical loading is applied to the PDMS substrate and sample through a house-made mechanical stretcher which implements the uniaxial mechanical stretching as shown in Figure S3 of the supplementary material. Although the ultimate tensile strains of a MoS$_2$ monolayer predicted by the first principle and MD simulations can be much higher than 10%[25], we find that fracture occurs in a MoS$_2$ monolayer when the tensile strain is around 8% in experiments as shown in the supplementary materials of Fig. S2. Therefore, in this study, we consider moderate strains ranging from 0% to 5.7% to avoid possible damage induced by tensile strains.

The thermal conductivity ($\kappa$) and interfacial thermal conductance per unit area ($g$), are obtained using the opto-thermal Raman approach. The scheme of the experimental setup is shown in Figure 1a. Figure 1b presents the optical image of the MoS$_2$ monolayer transferred onto the PDMS substrate prior to the application of strain. The total length of the monolayer in the strain direction is measured as 17.04 µm. Figure 1c shows an example of the MoS$_2$ monolayer after applying tensile strain by stretching the PDMS substrate. Now, the total length is around 18.11 µm. The successful strain transfer from the PDMS substrate to the overlying MoS$_2$ can be identified by comparing the lengths and the engineering strain can be calculated as 6.3%.

Figure 2a shows the Raman spectra of the unstrained MoS$_2$ monolayer at different temperatures. At 300 K, we observed the $E_{2g}^1$ vibration mode near 385 cm$^{-1}$ and the A$_{1g}$ vibration mode near 404 cm$^{-1}$. The frequency difference between the two vibration modes can be employed as the indicator of the thickness since the $E_{2g}^1$ vibration softens (red shifts), while the A$_{1g}$ vibration stiffens (blue shifts) with the increase of the MoS$_2$ thickness[26]. Atomic force microscopy (AFM) has also been used to characterize MoS$_2$'s thickness to reconfirm its thickness (Figure S4). This is essentially results from the change of the long-range Coulombic interlayer interactions for the $E_{2g}^1$ vibration and the decrease of the force constant due to the weakening of the interlayer Van der Waals force for the A$_{1g}$ vibration[27]. Accordingly, the frequency difference of a MoS$_2$ monolayer is around 19 cm$^{-1}$ and a MoS$_2$ bilayer is higher than 21 cm$^{-1}$. The frequency difference of our MoS$_2$ sample at 300 K is around 19.3 cm$^{-1}$, which can be confirmed as a monolayer.

The $E_{2g}^1$ mode is selected to determine $\chi_P$ and $\chi_T$ for the following two reasons. Firstly, we find that the intensity of the in-plane vibration mode $E_{2g}^1$ is much more significant than that of the out-of-plane vibration mode A$_{1g}$ as showing in Figure 2a. Additionally, it is reported that the $E_{2g}^1$ vibration is more sensitive to mechanical strain as compared with that of the A$_{1g}$ vibration[28]. This is because the $E_{2g}^1$ mode results from the opposite vibration of the two S atoms with respect to the Mo atom in the basal plane, while the A$_{1g}$ mode involves only the out-of-plane vibration of the S atoms in opposite directions. The tensile strain is applied to the basal plane of MoS$_2$ and thus directly influences the in-plane vibration of the Mo and S atoms. The A$_{1g}$ mode mainly depends on the stiffness of the out-of-plane spring-constants, which is relatively unaffected by the in-plane strain. Hence, the in-plane tensile strain principally influences the $E_{2g}^1$ vibration mode of the MoS$_2$ monolayer. The similar results are also reported on a mechanical exfoliated MoS$_2$ monolayer as Wang et al. reported that the $E_{2g}^1$ mode of the MoS$_2$ monolayer shows a visible red shift when applying uniaxial strain up to 3.6%, while the A$_{1g}$ mode keeps unchanged[24].

It is observed that the frequency of the $E_{2g}^1$ vibration model exhibits a red-shift proportional to the temperature as shown in Fig 2a. This temperature-dependent change in the frequency can be attributed to the thermal expansion of the lattice and/or the anharmonic temperature contribution induced by the coupling between phonons having different momentum and band index[29]. The Raman spectra of the MoS$_2$ monolayer subjected to different strains are all showing the similar trend. Figure 2b and 2d exhibit the change in the peak position of the $E_{2g}^1$ mode of the MoS$_2$ monolayer without strain and under the tensile strain of 6.3% as a function of temperature, respectively. The measurements were conducted on three MoS$_2$ samples and the measurement on each sample was repeated three times. The first-order temperature coefficient, $\chi_T$, is then extracted by fitting the data of the peak position versus the temperature via linear regression: $\omega(T) = \omega_0 + \chi_T T$, where $\omega(T)$ and $\omega_0$ are the frequencies of the $E_{2g}^1$ vibration mode at temperature $T$ and absolute zero, respectively. Figure 2c and 2e show the changes in the peak position induced by the Raman laser for the MoS$_2$ monolayer without strain and under the tensile strain of 6.3% as a function of the absorbed laser power, respectively. Likewise, the laser power-dependent Raman peak shift rates measured by both of the 50× and 100× lens are extracted by fitting the peak position versus the absorbed laser power via: $\omega(q_0) = \omega_0 + \chi_P q_0$, where $\omega(q_0)$ and $\omega_0$ are the frequencies of the $E_{2g}^1$ vibration mode subjected to laser heating at the power of $q_0$ and without laser heating

The thermal conductivity and the interfacial thermal conductance of the MoS$_2$ monolayer at different tensile strains are extracted and summarized in Table 1 along with the first-order temperature coefficients and laser power-dependent Raman peak shift rates. Prior to tensile strain, the obtained thermal conductivity value is 34.3±4.6 W/(m·K) for the $MoS_2$ monolayer. While most of the previous experimental work on $MoS_2$'s thermal conductivities were on the suspended samples[22-23, 30-31, 32], this value is in agreement with the only one work on supported $MoS_2$ with a thermal conductivity of 55±20 W/(m·K) [22].

Figure 3 shows the normalized thermal conductivity of the $MoS_2$ monolayer measured via the optothermal Raman method as a function of tensile strain, along with the normalized thermal conductivity predictions from former simulations. Our results demonstrate a significant descending trend for the thermal conductivity of the $MoS_2$ monolayer versus tensile strain. More specifically, a 6.3% tensile strain reduces the thermal conductivity by around 62%. Such observation is in agreement with the conclusion drew by the computational work of Jiang et. al[14], Ding et. al[15], Zhu et. al[17], and Xiang et. al[16], which predicts that the in-plane thermal conductivity of a $MoS_2$ monolayer can be heavily suppressed by tensile strains. While the thermal conductivity of $MoS_2$ predicted by the computational work is shown to monotonically decrease with tensile strain, our results show that the thermal conductivity of the sample under the strain of 1.6% is slightly higher than that of the sample prior to the strain which trend is in agreement with the computational work of thermal conductivity of graphene with strain [33]. Same as the "twist" in graphene without strains, $MoS_2$ transferred on PDMS substrates is non-flat and this increases the interfacial phonon scattering and thus decreases the thermal transport. A small applied strain is balanced to minimize the interfacial torsion of $MoS_2$, and a higher thermal conductivity is observed at 0% - 1.6% strain. When the applied strain is beyond the 1.6% strain point, the Mo-S bonds start being stretched, which results in a decrease in thermal conductivity due to softened phonon modes. The similar decreasing effect has also been reported and studied in compressed graphene and $MoS_2$ monolayers via MD simulations[15, 34]. With the application of tensile strain, our $MoS_2$ monolayer is gradually stretched. Meanwhile, the out-of-plane deformation is smoothed by the in-plane tensile strain, weakening the phonon scattering, which increases the thermal conductivity of the $MoS_2$ monolayer.

EXPERIMENTAL SECTION

CVD-grown $MoS_2$ monolayer is employed since continuous $MoS_2$ monolayer can be produced at micro or even centimeter scale [35-36]. To transfer the $MoS_2$ samples on the PDMS substrate, a thin layer of PMMA is coated on the CVD $MoS_2/SiO_2/Si$ stack. The assembly remains at room temperature to remove the solvent of the PMMA. Then, the $PMMA/MoS_2/SiO_2/Si$ stack is placed on the potassium hydroxide (KOH, 15% w/v) solution in a petri dish. Before the stack is placed in the petri dish, the periphery of the PMMA has been partially scratched using a blade so that the KOH solution can penetrate at the interface of the PMMA and $SiO_2/Si$. From this process, the KOH solution can easily dissolve the thin $SiO_2$ layer, leading to the separation between the $PMMA/MoS_2$ stack and the Si substrate. Afterward, the $PMMA/MoS_2$ stack floating on the KOH solution is rinsed using DI water for three times to remove the possible KOH contaminants. After rinsing, the $PMMA/MoS_2$ stack is transferred onto the PDMS substrate. The transferred $PMMA/MoS_2/PDMS$ stack is dried in a desiccator overnight at room temperature for removing water molecules and for a better adhesion between the $PMMA/MoS_2$ stack and the PDMS substrate. To remove the PMMA, the $PMMA/MoS_2/PDMS$ stack is immersed in an acetone solution for 30 minutes at 45°C.

CONCLUSION

We have successfully manufactured a $MoS_2$ monolayer onto a flexible substrate and conducted optothermal Raman spectroscopy to investigate the thermal conductivity of the $MoS_2$ monolayer subjected to tunable uniaxial tensile strains. Our results reveal that the in-plane thermal conductivity of MoS$_2$ monolayer dramatically decreases with tensile strain. Regarding the PDMS supported MoS$_2$ monolayer, it is found that its in-plane thermal conductivity drops by 62% under a 6.3% uniaxial tensile strain. Our result support the previous DFT and MD simulation predictions that tensile strains can significantly suppress the in-plane thermal conductivity of a MoS$_2$ monolayer. This work verifies that the mechanical strain can be used to tune the figure of merit of 2D MoS$_2$ monolayer, which is crucial to the exploration of both fundamental physics and high-performance wearable and flexible devices.

FIGURES.

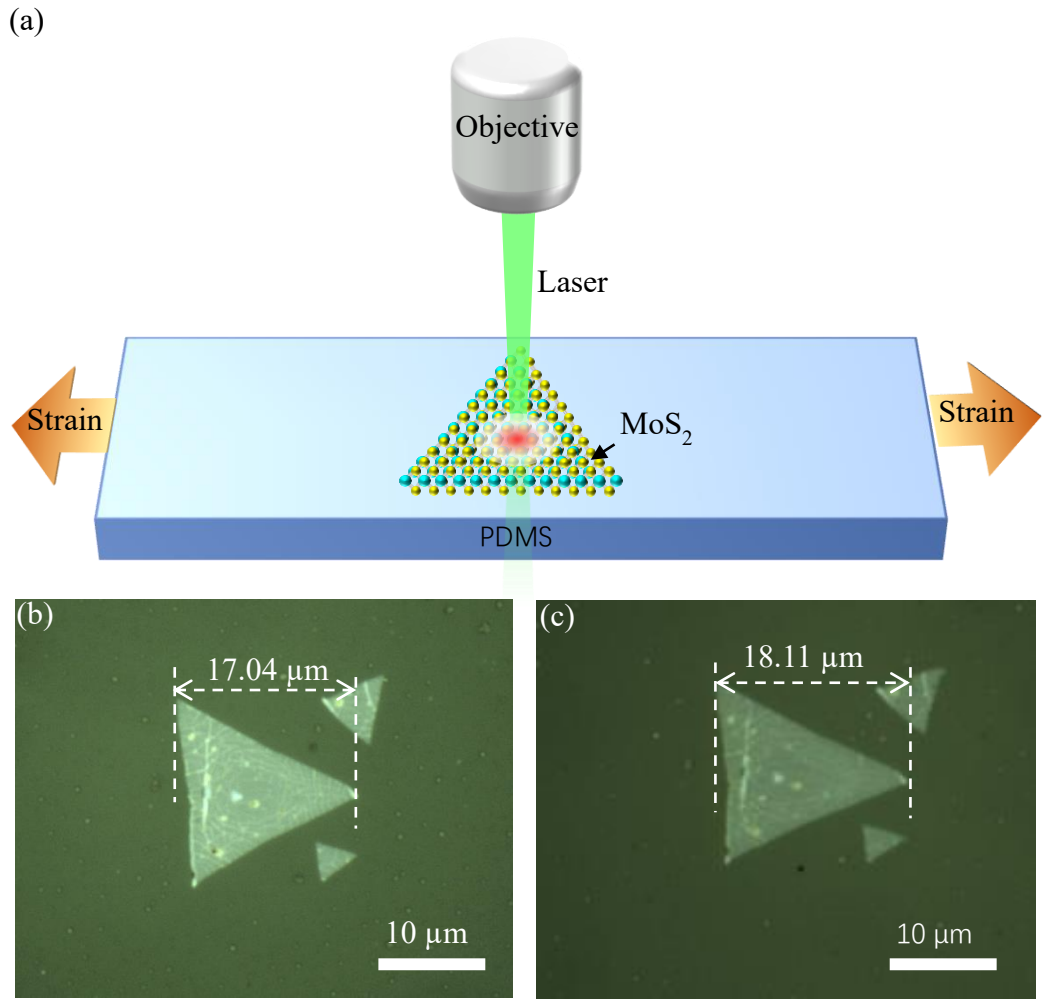

**Figure 1**. (a) Schematic of the optothermal Raman experiment setup for the strained $MoS_2$ monolayer. (b) Morphology of the $MoS_2$ monolayer prior to tensile strain captured using an optical microscope. (c) Morphology of the $MoS_2$ monolayer subjected to the tensile strain of 6.3% captured using the optical microscope.

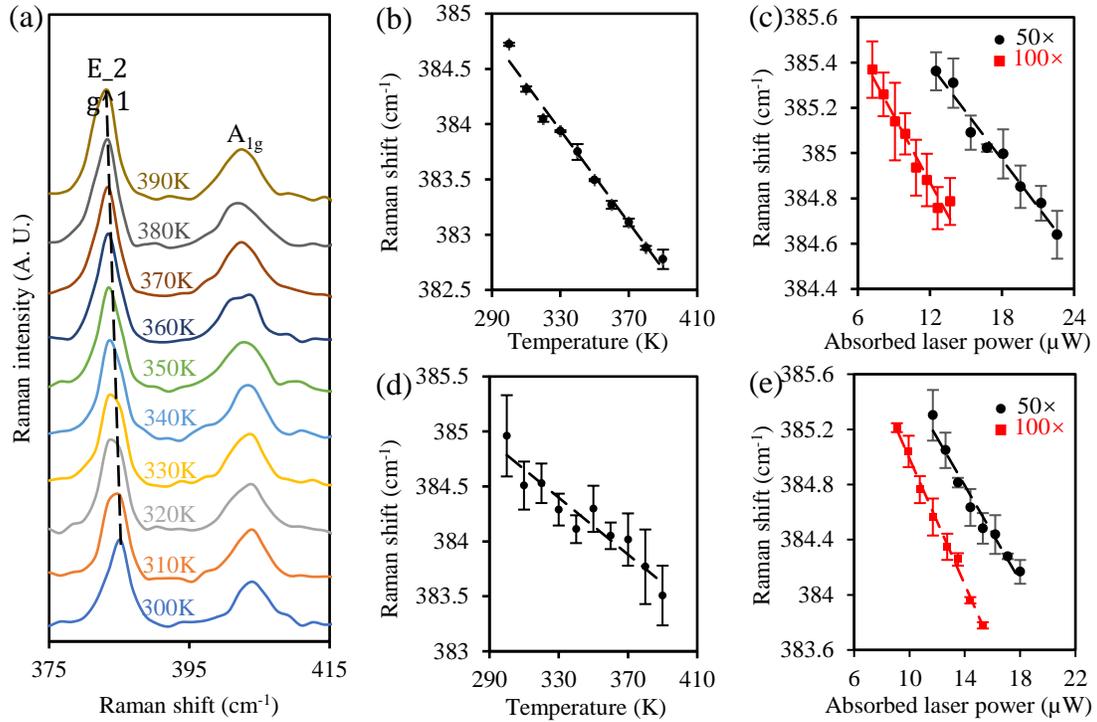

**Figure 2**. (a) Evolution of the Raman spectra of the MoS$_2$ monolayer on the PDMS substrate at different temperatures. The dashed line is the guide of the peak centers. (b) The temperature-dependent $E_{2g}^1$ Raman peak shift measured on the MoS$_2$ monolayer prior to tensile strain. (c) Power-dependent $E_{2g}^1$ Raman peak shift measured using different optical lens on the MoS$_2$ monolayer prior to tensile strain. (d) The temperature-dependent $E_{2g}^1$ Raman peak shift measured on the MoS$_2$ monolayer under the tensile strain of 6.3%. (e) Power-dependent $E_{2g}^1$ Raman peak shift measured using different optical lens on the MoS$_2$ monolayer under the tensile strain of 6.3%.

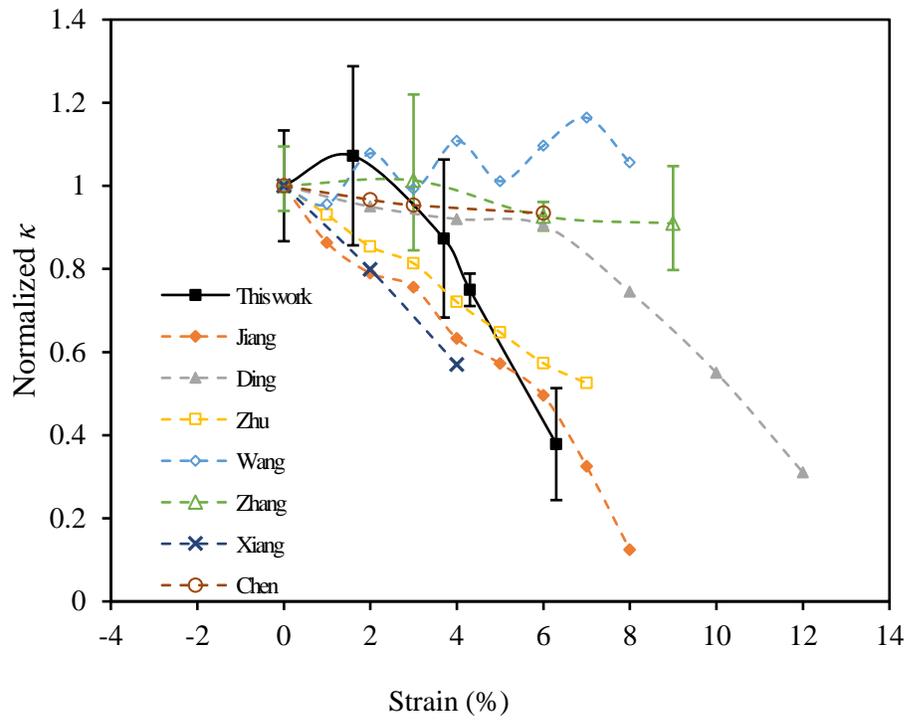

**Figure 3**. Normalized thermal conductivity of the $MoS_2$ monolayer as a function of tensile strain and comparison with computational model predictions[14-19, 37].

TABLES.

**Table 1**. First-order temperature coefficients, power shift rates, in-plane thermal conductivities, and interfacial thermal conductance of the CVD MoS$_2$ monolayer under different uniaxial tensile strains

| Strain (%) | $\chi_T$ (cm$^{-1}$/K) | $\chi_P$ (cm$^{-1}$/μW) | | k (W/(m·K)) | g (MW/(m$^2$·K)) |
|---|---|---|---|---|---|
| | | 50× | 100× | | |
| 0 | -0.0209±0.0008 | -0.0699±0.0046 | -0.0976±0.0078 | 34.3±4.6 | 0.19±0.01 |
| 1.6 | -0.0193±0.0009 | -0.0901±0.0058 | -0.1192±0.0107 | 36.8±7.4 | 0.10±0.02 |
| 3.7 | -0.0161±0.0008 | -0.1320±0.0125 | -0.1657±0.0096 | 29.9±6.5 | 0.04±0.02 |
| 4.3 | -0.0155±0.0006 | -0.1587±0.0067 | -0.1971±0.0068 | 25.7±1.3 | 0.02±0.01 |
| 6.3 | -0.0129±0.0015 | -0.1730±0.0136 | -0.2288±0.0075 | 12.9±4.62 | 0.03±0.01 |

**Data Availability**

The datasets used and analysed during the current study available from the corresponding author on reasonable request.

**Acknowledgements**

We acknowledge the National Science Foundation to support our work (CAREER Award (Grant CBET-2145417) and LEAPS Award (Grant DMR-2137883)).

**Author Contributions**

X.Z. contributed to the conceptualization and the methodology. J.L. and X.Z. contributed to the investigation. M.F. and E.Y. contributed to the material synthesis. X.Z. and E.Y. contributed to the supervision. J.L. contributed to writing the original draft. X.Z. and J.L. contributed to writing the review and editing. All authors have given approval to the final version of the manuscript.

**Declarations**

**Competing Interests**

The authors declare no competing interests.

**Additional Information**

**Supplementary Information** The online version contains supplementary material. (Laser spot size characterization, captured tensile strain, mechanical stretcher, AFM characterization of thickness.)

**Corresponding Author**

*Xian Zhang. Email: xzhang4@stevens.edu

**Funding Sources**

National Science Foundation CAREER Award (Grant CBET-2145417) and LEAPS Award (Grant DMR-2137883).